\pdfoutput=1
\documentclass[%
 aip,
 amsmath,amssymb,
 reprint,%
]{revtex4-1}

\usepackage{graphicx}
\usepackage{dcolumn}
\usepackage{bm}

\usepackage[utf8]{inputenc}
\usepackage[T1]{fontenc}
\usepackage{mathptmx}
\usepackage{etoolbox}
\usepackage{wasysym}
\usepackage{hyperref}
\usepackage{breakurl}
\usepackage[english]{babel}
\renewcommand{\selectlanguage}[1]{}
\usepackage{url}

\usepackage{breakurl}

\makeatletter
\def\@email#1#2{%
 \endgroup
 \patchcmd{\titleblock@produce}
  {\frontmatter@RRAPformat}
  {\frontmatter@RRAPformat{\produce@RRAP{*#1\href{mailto:#2}{#2}}}\frontmatter@RRAPformat}
  {}{}
}%
\makeatother

\begin{document}

\preprint{AIP/123-QED}

\title{Characterization of Spatiotemporal Overlap of Femtosecond Lasers and Electron Beam With Ce:YAG Screens
}

\author{Kevin Eckrosh}
\affiliation{Department of Physics, Arizona State University, Tempe AZ}\email{Keckrosh@asu.edu}
\author{Sean Tilton}%
\affiliation{Department of Physics, Arizona State University, Tempe AZ}%

\author{Lucas Malin}
\affiliation{%
Department of Physics, Arizona State University, Tempe AZ
}%

\author{Taryn Brown}
\affiliation{%
Department of Materials Science and Engineering, Arizona State University, Tempe AZ
}%

\author{Alan Dupre}
\affiliation{%
Department of Physics, Arizona State University, Tempe AZ
}%

\author{Antonella Semaan}
\affiliation{%
Biodesign Institute, Arizona State University, Tempe AZ
}%

\author{Alex Gardeck}
\affiliation{%
Biodesign Institute, Arizona State University, Tempe AZ
}%

\author{Gregory Babic}
\affiliation{%
Department of Physics, Arizona State University, Tempe AZ
}
\affiliation{Biodesign Institute, Arizona State University, Tempe AZ}

\author{Hyung Seo Lee}
\affiliation{%
Department of Physics, Arizona State University, Tempe AZ
}
\affiliation{Biodesign Institute, Arizona State University, Tempe AZ}

\author{Henrik Loos}
\affiliation{Biodesign Institute, Arizona State University, Tempe AZ}

\author{Mukhtar Hussein}
\affiliation{%
Biodesign Institute, Arizona State University, Tempe AZ
}
\affiliation{Biodesign Institute, Arizona State University, Tempe AZ}

\author{Arvinder Sandhu}
\affiliation{%
Department of Physics, University of Arizona, Tuscon AZ
}%

\author{William S. Graves}
\affiliation{%
College of Integrative Arts and Sciences, Arizona State University, Tempe AZ
}%
\affiliation{%
Center for Applied Structural Discovery, Biodesign Institute, Arizona State University, Tempe AZ
}%

\author{Mark R. Holl}
\affiliation{%
Biodesign Institute, Arizona State University, Tempe AZ
}%

\author{Samuel W. Teitelbaum}
\affiliation{%
Department of Physics, Arizona State University, Tempe AZ
}%
\affiliation{%
Center for Applied Structural Discovery, Biodesign Institute, Arizona State University, Tempe AZ
}%



\date{\today}

\begin{abstract}
Interactions between short laser pulses and electron bunches determine a wide range of accelerator applications. Finding spatiotemporal overlap between few-micron-sized optical and electron beams is critical, yet there are few routine diagnostics for this purpose. We present a method for achieving spatiotemporal overlap between a picosecond laser pulse and a relativistic sub-ps electron bunch. The method uses the transient change in optical transmission of a Ce:YAG screen upon irradiation with a short electron bunch to co-time the electron and laser beams. We demonstrate and quantify the performance of this method using an inverse Compton source comprised of a 30 MeV electron beam from an X-band linac focused to a 10 $\mu$m spot, overlapped with a joule-class picosecond Yb:YAG laser system. This method is applicable to electron beams with few-microjoule bunch energies, and uses standard scintillator screens common in electron accelerators. 
\end{abstract}

\maketitle


%


\section{\label{sec:I}Introduction}

Electron accelerators are used in applications ranging from bright x-ray production\citep{graves_compact_2013, hartemann_compton_2007, krafft_compton_2010, albert_applications_2016, matlis_ultrafast_2010, tajima_wakefield_2020} to medical therapies\citep{jacquet_radiation_2015}. The integration of ultrafast lasers with electron beams can improve the performance and expand the application scope of electron accelerators\cite{huang_measurements_2010}, enabling compact x-ray sources\cite{graves_compact_2013}, attosecond x-ray pulses\cite{spampinati_laser_2014}, and compact high-energy particle beams\cite{esarey_physics_2009}. Illustrative applications of laser-electron beam interactions include plasma wakefield accelerators\cite{esarey_physics_2009}, inverse Compton scattering\cite{fuchs_laser-driven_2009}, and laser heaters\cite{huang_measurements_2010}.

Fully leveraging the opportunities of modern high-power laser technology requires achieving precise and stable spatiotemporal overlap between a relativistic electron beam and a high power laser. Therefore, development of an approach that can independently diagnose both spatial and temporal overlap between the electron and laser beam is critical to rapid and productive accelerator alignment. Ideally, the instrumentation should be low-cost and accessible in a form factor for use in compact, relatively low-energy, low-charge accelerators where electro-optic sampling\cite{cavalieri_clocking_2005} may not have the sensitivity due to the small Coulomb field of the electron bunch.

Here, we demonstrate a compact diagnostic using transient absorption in Cerium-doped YAG (Ce:YAG) that can address the requirement for obtaining spatial and temporal overlap of a laser and a relativistic electron beam. We show that this diagnostic is capable of quantifying spatial overlap to few-micron precision (on the order of the size of the laser beam) and temporal overlap to sub-picosecond precision. The diagnostic requires nanojoules of laser energy and few-pC bunch charges at 10s of MeV beam energies. We demonstrate one application of this method, measuring the long-term spatiotemporal overlap stability of the electron and laser beam in an inverse Compton scattering (ICS) x-ray source operating at 1 kHz\cite{graves_compact_2013}. The instrumentation required for this diagnostic is compact and low-cost, as it only requires a Ce:YAG screen on a translation stage and near-IR and visible cameras.

\subsection{\label{sec:I.A}Holey YAG For Spatial Overlap}
\begin{figure*}
  \includegraphics[width=\textwidth]{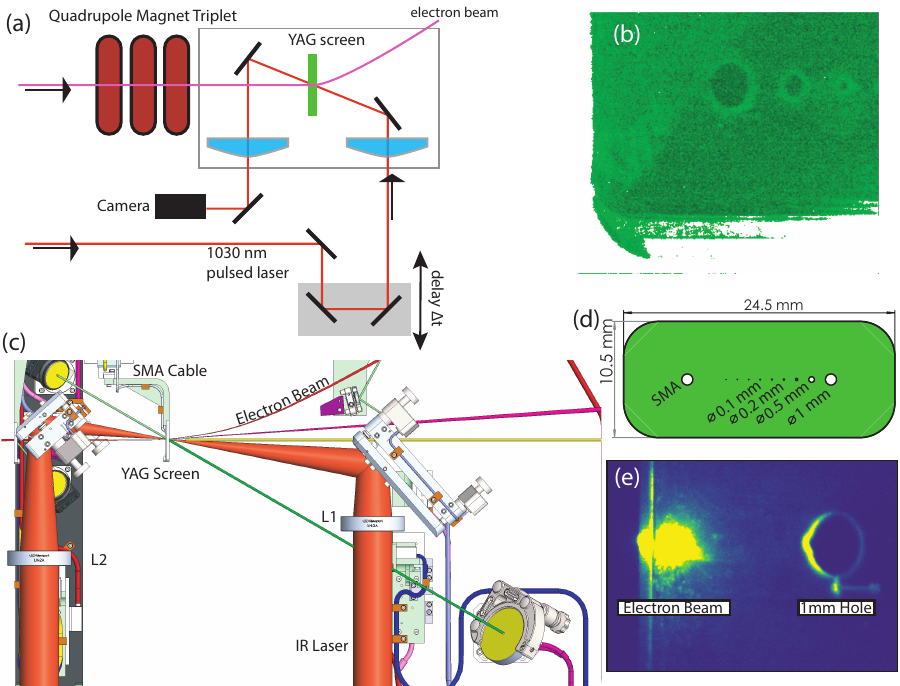}
  \caption{(a) Experimental setup used for the holey YAG and transient YAG approaches. The IR beam transmitted through the YAG screen is imaged onto a camera 692.9 mm away from the screen. Experimental chamber is outlined in gray. (b) Image of part of the holey YAG screen. The largest hole shown is 1mm in diameter. (c) CAD model of key components of the experimental chamber used for electron and laser beam interaction. The SMA is attached to a hole in the YAG shown in (d). The green line indicates the line of sight to the camera which produces the image shown in (e). (d) Layout of holes in the holey YAG screen. The hole diameters decrease from right to left. The hole used for the SMA cable test is on the far left, and is \diameter 1mm. (e) Fluorescence image of the electron beam on the holey YAG screen as seen by the camera. The \diameter 1 mm hole in the holey YAG is shown on the right for comparison.  }
  \label{fig:Setup}
\end{figure*}

A laser-machined Ce:YAG screen is placed at normal incidence along the propagation axis of the electron beam, and the electron-irradiated region of the screen is probed with a counter-propagating IR laser as shown in Fig 1(a). A camera (Allied Vision Mako series) is placed on the roof of the chamber to image the fluorescence of the YAG screen due to the incident electrons and laser irradiation. The YAG screen has a series of micro-machined holes drilled into it (Fig \ref{fig:Setup})(b) and Fig \ref{fig:Setup})(d). Nine of these holes (\diameter 1 mm, \diameter 0.5 mm, \diameter 0.2 mm,\diameter 0.1 mm, \diameter 0.07 mm, \diameter 0.05 mm, \diameter 0.03 mm, \diameter 0.02 mm, \diameter 0.01 mm) are used for estimating the size of the electron beam, and roughly aligning the laser to the nominal interaction point. An additional \diameter 1 mm hole features the stripped end of an SMA cable placed parallel to the plane of the YAG screen, with its center extending to the center of the hole. This hole is for the timing diagnostic described in \ref{sec:I.B}.


\subsection{\label{sec:I.B}Coarse Temporal Overlap}

Before using a mechanical delay stage to scan for timing, it is necessary to ensure temporal overlap can be found within the temporal scan range of the stage (approximately 300 ps). We accomplish this by using real-time electronic signals from the laser and electron beam that can be read out on an oscilloscope. Fig \ref{fig:Setup}(d) shows a large hole on the left of the YAG for one end of an SMA cable to be attached to. This end of the SMA is stripped, such that a current will be induced when an electron bunch passes through the hole. Similarly, laser pulses will induce a current via ablation and ionization when they strike the wire. Each of these signals are read on an oscilloscope, which displays voltage spikes corresponding to the timing of each beam relative to an external trigger. Any offset is corrected by adjusting only the delay time of the laser. The resolution of this method is limited by the bandwidth of the oscilloscope (WavePro 404HD, 4 GHz bandwidth), which limits our resolution of the center of the rise to approximately 25 ps. The timing of the laser and electron beam is achieved with a combination of electronic triggering and mechanical delays. Adjusting the electronic triggering of the laser brings the beams within a single oscillator period (13.8 ns) by adjusting which oscillator pulse is amplified by the 1 kHz laser amplifier. Coarse delay control with sub-ns precision is achieved with a long-pass delay stage (Newmark Systems NLS4-24-12-NC) located between the laser amplifier and grating compressor. This stage adjustment is sufficient to bring the timing within the travel range of the 100 mm long fine-control delay stage (Newport M-ILS200BPP). 

\subsection{\label{sec:I.C}Transient YAG Approach}

The transient absorption induced in the YAG screen allows for higher precision measurements and fine tuning of the spatiotemporal overlap that cannot be achieved with the previous two methods. Each pulse of the electron beam induces a spatially localized and temporary decrease in the transmission of IR through the YAG screen. The laser beam transmitted through the YAG screen is loosely focused onto a camera which monitors the relative transmitted intensity as the spatial and temporal positions of the IR and electron pulses are adjusted.  This approach is analogous to similar approaches developed for monitoring fine timing between X-ray and optical pulses\cite{sato_simple_2019,diez_self-referenced_2021,lemke_femtosecond_2013,nakajima_software_2018}.

Scanning the laser spatially across the screen produces changes in intensity that follow an inverted Gaussian curve, whose center corresponds to the position of maximum spatial overlap. The process is repeated for both horizontal and vertical axes by translating the laser focusing lens L1.  At each delay or spatial overlap step, reference (no electron beam) images are acquired by adjusting the electron beam accelerator phase by $180^{\circ}$ to correct for long-term drifts in the laser intensity.

\section{\label{sec:II}Results}

The electron beam used in this study is produced by a photoinjector\cite{graves_design_2017} followed by three accelerating cavities that operate in the X-band at an RF frequency of 9300 MHz driven by two Scandinova klystrons\cite{cook_design_2022}. The accelerator operates at a 1 kHz repetition rate. A series of three quadrupole magnets focus the electron beam to a 10$\mu$m RMS spot at the interaction point. A $30^{\circ}$ bend magnet just after the laser-electron beam interaction directs the beam to a dump  (Fig. \ref{fig:Setup}(a)).


The photocathode laser is a Yb:YAG Light Conversion Pharos laser producing 10.5 uJ of 4th harmonic (275 nm) incident light. The total photo-charge of the electron beam was 28 pC (quantum efficiency of 1.84e-5). The final electron energy after acceleration is 31.8 MeV.  The UV beam is shaped to a flat-top profile on the cathode with a diameter of 800 microns. 

The probe laser is an attenuated thin-disk Yb:YAG laser (Trumpf DIRA-200)\cite{hong_highly-stable_2018,nubbemeyer_1_2017} which is attenuated to 6 uJ of pulse energy for this study. It has a pulse duration of 1.5 ps FWHM. A lens of focal length $f=228$ mm focuses the laser to a spot size of 10 $\mu$m $1/e^2$ diameter at the interaction point.

\subsection{\label{sec:II.A}Coarse Overlap Scans}
\begin{figure}
  \includegraphics[width=\columnwidth]{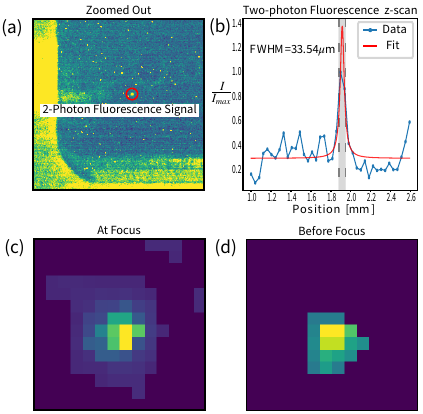}
  \caption{Images of the two-photon fluorescence of the IR beam on the YAG screen as seen by the camera. (a) The full image seen from the camera. The intensity of the two-photon fluorescence signal reaches a maximum when the YAG screen is placed at the focus of the IR beam. (b) The YAG screen is scanned through the direction of propagation of the IR beam. The intensity of the two-photon fluorescence signal is recorded along the scan and fit to a Lorentzian to find the position of the focus. (c and d) Zoomed in images of the two-photon fluorescence signal showing the difference in intensity at the focus (c) and outside of the focus (d).}
  \label{fig:Overlap}
\end{figure}

The electron and laser beams cross at a finite angle (approximately 11 degrees in a counterpropagating geometry), as shown in Fig \ref{fig:Setup}(a) and Fig \ref{fig:Setup}(c). Thus, the two beams can only be overlapped across a small region in space. This should occur at the focus of the laser and electron beams in order to maximize the efficiency of the Inverse Compton Scattering process. 

The first step to achieving spatiotemporal overlap is to place the focus of the electron and laser beams in the plane of the Ce:YAG screen at the interaction point. For the laser, this is achieved by scanning the YAG screen along the propagation axis of the laser to find the focal position of the laser. We use the intensity of the two-photon fluorescence on the YAG as our diagnostic for focus quality; the position that maximizes the intensity of the two-photon fluorescence minimizes the beam diameter on the YAG screen. Fig \ref{fig:Overlap}(a) shows a typical image of the YAG screen with the fluorescence signal. The intensity within the region of interest (ROI) near the laser beam as a function of YAG position is shown in Fig \ref{fig:Overlap}(b). Fig \ref{fig:Overlap}(c) and Fig \ref{fig:Overlap}(d) show zoomed in images of the fluorescence spot at different levels of focus. The spatial resolution of the imaging camera is insufficient to directly measure the spot size of the laser beam from the fluorescence spot. The beam spot size (estimated at 10 $\mu$m based on the focal length of the lens and the beam's $M^2$) is smaller than the spatial resolution of the imaging system (approximately 50 $\mu$m).

Now that the optimal focal plane of the laser is found, the lens focusing axis and YAG screen are moved together such that the YAG is at the focus of the electron beam. Alternatively, the electron focusing optics can be adjusted to achieve the desired interaction position as given by the position of the Ce:YAG screen. In all cases, this method defines the spatiotemporal interaction point as being in the plane of the Ce:YAG screen. 

\subsubsection{\label{sec:II.A.1}Holey YAG Scan}

With the focus of the laser now in the plane of the YAG screen, the next priority is to find coarse spatial overlap using the holey YAG method described in section \ref{sec:I.A}. The general idea of the holey YAG approach is to use the holes as apertures for each beam to be centered upon. Since we know the spacing between each hole, we can vary the size of our aperture by translating the YAG screen 1 mm plus the sum of the radii of the current hole and adjacent hole. Therefore, the entire array of holes can be used together as an effective variable aperture to adjust the precision of this method. It is generally simpler to move the transverse positions of the YAG screen and the laser beam rather than the that of the electron beam, and thus, we make adjustments only on the former two throughout this method. 

The first step is to move the YAG screen such that the largest of its holes is centered around the electron beam. The laser is blocked so that any fluorescence around the edge of the holes is solely due to the electron beam. We know we are centered when this fluorescence is symmetric around the edges of the hole. Then, the electron beam is turned off, and the laser is unblocked. We center the laser beam in a similar manner, only this time, the YAG screen is held at a fixed position, and the laser position is adjusted by translating a lens upstream. Once centered, we block the laser again, turn on the electron beam, and repeat the process with progressively smaller holes each iteration. 

For the larger holes, the laser spot is too small to see two-photon fluorescence on the edges. In these cases, we center the laser by scanning the YAG screen across the hole and observing the clipping of the laser beam on a camera that images the laser beam profile. Small adjustments are made with the focusing lens translation stage. The control value for the corresponding axis on the lens' translation stage is recorded, and their superposition is taken as the center. The holey YAG approach is not precise enough to single-handedly achieve the level of spatial overlap we need, but it is necessary to perform before moving on to the higher precision transient YAG method. This method will typically be able to position the laser beam within 5 $\mu$m of a target position. 
\subsubsection{\label{sec:II.A.2}Coarse Temporal Scan (SMA Cable Test)}

\begin{figure}[ht]
  \includegraphics[width=\columnwidth]{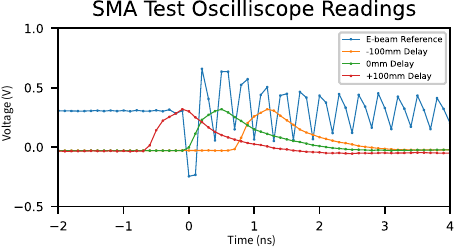}
  \caption{Compact diagnostic for coarse timing of a laser and electron beam with a bare SMA cable. The oscillatory signal is the SMA cable pickup when the electron beam passes near the SMA cable. The electron beam arrival time is given by the early negative spike. Three laser arrivals are shown, negative extreme, center and positive extreme of a 200 mm long optical delay line. }
  \label{fig:SMA}
\end{figure}

With both beams approximately overlapped in space, we can begin the search for temporal overlap. Since the initial relative temporal positions of each pulse could be anywhere in the 1kHz cycle, we must first employ the coarse timing approach described earlier. This method determines the relative arrival of the electron and laser beam to sub-nanosecond precision, and will sync the arrival of each pulse close enough such that their relative times can seen by the few-picosecond precision transient YAG method, within the timing window of a mechanical delay stage. 

A 1 mm diameter hole in the YAG screen shown in Fig \ref{fig:Setup}(d) is outfitted with a stripped SMA with one end hooked up to an oscilloscope as described in section \ref{sec:I.B}. First, the electron beam is passed near the bare SMA cable.  The transient Coulomb field of the bunch generates an RF pickup on the SMA cable (Fig \ref{fig:SMA}). The signal's trace is saved on the oscilloscope to be used as a reference, and the electron beam is turned off.

The laser beam is then passed through the same hole such that it is incident on the bare SMA cable. Raising the power of the laser produces a transient signal due to ionization at the SMA cable tip. This voltage spike is compared with the timing of the electron beam reference signal, as shown in Fig \ref{fig:SMA}. 

This method cannot be used alone to provide the level of temporal precision we require. Similarly to the scans of section \ref{sec:I.A}, it must be performed before moving on to higher precision methods. This is because the laser is seeded off the same 72 MHz laser oscillator as the photoinjector, and is therefore intrinsically locked to sub-picosecond precision, up to a factor of an integer multiple of the oscillator repetition rate of 13.8 ns. The coarse temporal scans allow us to determine which 13.8 ns bucket to select.

\subsection{\label{sec:II.B}Fine Overlap Scans}

\begin{figure}
  \includegraphics[width=\columnwidth]{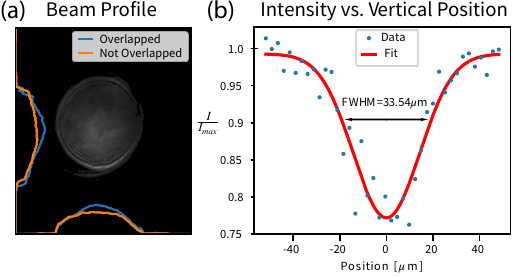}
  \caption{ Using the YAG screen to achieve spatial overlap of the electron and laser beams. (a) Image of the laser beam imaged on the camera shown in figure \ref{fig:Setup}(a). Blue and orange curves are vertical and horizontal sums with the laser beam at the beginning and middle of the spatial overlap scan. (b) Total intensity of the transmitted IR beam as a function of the focusing lens vertical (y) position. The decrease in the middle indicates there is spatial overlap between the electron and laser beams.}
  \label{fig:Spatial_scan}
\end{figure}

We now have approximate spatiotemporal overlap between the electron beam and laser to use the higher precision transient YAG method described in section \ref{sec:I.C}. A section of the YAG screen with no holes is moved to the intersection of the counterpropagating beams. A camera views the transmitted laser beam through the YAG screen as shown in Fig. \ref{fig:Setup}. 

When the electron beam arrives at the YAG screen, the electrons inelastically scatter off the carriers in the YAG via interaction with their Coulomb field, generating a population of free carriers. These free carriers subsequently absorb any incoming infrared radiation, reducing the transmission of the YAG screen. Therefore, monitoring the transmitted intensity of the IR beam provides a diagnostic to independently measure the spatial and temporal overlap between the two beams. 

Optimizing overlap with this method begins by scanning the IR beam across the electron beam fluorescence on the YAG screen. The camera takes a series of images of the laser transmission shown in Fig \ref{fig:Spatial_scan}(a). The total intensity of these images is plotted and fit to a Gaussian curve of the form 

\begin{equation}
    g(x) = A \cdot \exp \left( \frac{(x-x_0)^2}{2\sigma^2} \right)+B
\end{equation}
in Fig. \ref{fig:Spatial_scan}(b), with the center of the curve corresponding to the position of optimal overlap. The scan is performed over both the vertical and horizontal axes, and the laser beam is relocated to the optimal position found.

\begin{figure}
  \includegraphics[width=\columnwidth]{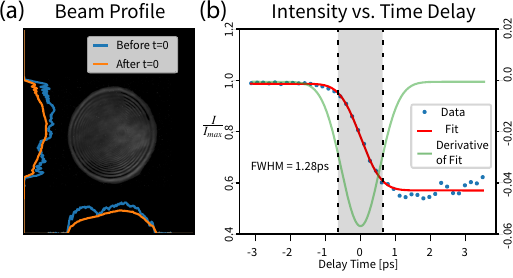}
  \caption{Using the YAG screen to achieve temporal overlap of the electron and laser beams. (a) Image of the laser beam imaged on the camera shown in figure \ref{fig:Setup}(a). Blue and orange curves are vertical and horizontal sums with the laser beam arriving before and after time $t=0$. (b) The total intensity of the transmitted IR beam as a function of delay time. The laser beam is initially early relative to the electron beam, and is therefore, unattenuated. As the laser pulse passes $t=0$ the transmitted intensity begins to drop due to the excitation from the electron beam. The position of temporal overlap is found by fitting the results to an error function and finding the extremum in its derivative. If the laser beam was initially late relative to the electron beam, the transmitted intensity would rise as it passes $t=0$. In this case, the extrema in the derivative of the error function would correspond to a maximum. }
  \label{fig:Temporal_scan}
\end{figure}

A similar method is employed to find temporal overlap, using a delay stage to scan the laser beam in time. Typically, we first measure the intensity change at the start and end of the delay stage position to confirm that temporal overlap is within the scan range of the stage. We then use a binary search method to narrow in on the overlap region until the stage position where the intensity changes is known to a few mm. We choose a start and end point for the scan that passes over the estimated time of overlap, $t_0$, found via the SMA cable test of section \ref{sec:II.A.2}. The laser intensity profile and results of this scan are shown in Fig\ref{fig:Temporal_scan}. These results are fit to an error function,
\begin{equation}
    f(t) = A \cdot \mathrm{erf} \left ( {\frac{t-t_0}{\sigma}} \right ) + B
\end{equation}
whose derivative is a Gaussian centered about $t_0$. Alternating spatial and temporal scans are repeated a few times, since each spatial scan can slightly alter $t_0$, and vice versa. 

\section{Resolution and Sensitivity}

\begin{figure*}[hbt!]
    \includegraphics[width =2\columnwidth]{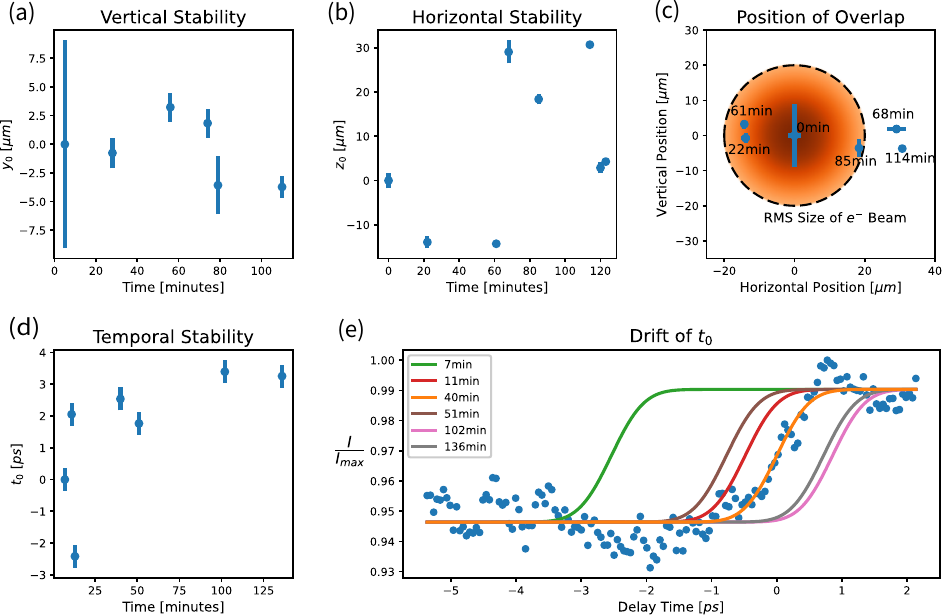}
    \caption{The updated spatial and temporal overlap positions taken periodically over 136 minutes of beam operations. (a, b) The vertical and horizontal drifts in the position of overlap. (c) The magnitude of spatial drift of the position of overlap shown relative to the RMS size of the electron beam. Note that the time corresponding to each data point is displayed. (d) The long-term drift of the optimal temporal position. (e) A reference temporal scan [40 minutes]is performed and fit to an error function to find temporal overlap. Resulting fits from subsequent scans are shown relative to the data and fit from the initial.}
    \label{fig:Drift}
\end{figure*}

We now discuss the expected temporal resolution of this timing method, the precision with which the electron and laser beam can be overlapped, and the minimum charge density required for reasonable sensitivity. Several factors determine the temporal width of the signal. These factors are the duration of the electron and laser pulses $\tau_e$ and $\tau_l$, the jitter between them $\tau_j$.  Furthermore, there is the response time of the Ce:YAG screen itself $\tau_{YAG}$, group velocity mismatch between the electron and laser beams in the YAG screen, and the effect of the finite thickness of the YAG screen because the beams counterpropogate.

The $11^{\circ}$ crossing angle of the electron and optical beams combined with the finite thickness of the YAG screen creates a correlation between the temporal and spatial (horizontal) overlap. If the laser arrives earlier, it will appear as if the laser has shifted positively in the horizontal axis. Likewise, if the laser arrives later, it will appear as if it has shifted negatively in the horizontal axis.  If we let the temporal and horizontal drifts be $\Delta t$ and $\Delta x$, respectively, then the correlation between the two axes is
\begin{equation}
    \sin{\theta} = \frac{\Delta x}{v \Delta t}
\end{equation}
Where $v$ is the group velocity of the laser in the YAG screen.

The finite overlap angle also smears the horizontal resolution of the spatial overlap scan by an amount $\sigma_x = L\tan(\theta)$, where $L$ is the thickness of the YAG screen. For our YAG screen, $L = 100 \mathrm{\mu m}$, adding $20 \mathrm{\mu m}$ (in quadrature) to the $1/e^2$ width of the horizontal scan. For our electron beam, the RMS spot size is 10 $\mu$m ($1/e^2$ diameter 40 $\mu$m) so this effect is negligible.

The overall temporal response function is a convolution of all the individual response functions from the factors above.  Under the assumption that all the above response functions are approximately Gaussian (other than the intrinsic overlap signal, which we assume is a step-like function), then the final response temporal width is simply the quadrature sum of the individual effects

\begin{equation}
    \tau_{tot} = \sqrt{\Big(  \tau_j^2 + \tau _l ^2 + \tau_e^2 + \tau_d^2 + \tau_{YAG}^2\Big)}
\end{equation}
In the data presented in Fig. \ref{fig:Temporal_scan} the laser pulse was measured to be 1.5 ps FWHM, the electron pulse duration is estimated to be 100 fs FWHM, the jitter is 282 fs FWHM, and the effect of the thickness of the YAG is 830 fs.  The response time of the YAG screen is negligible on these timescales of order of less than 200 fs \cite{sanchez-gonzalez_coincidence_2017}.  Added in quadrature, this results in an overall rise time of $\tau_{tot} = 1.7 ps$.

Finally, we discuss the sensitivity of this method. Here, a 25 MeV electron beam with 20 pC of charge is used, producing a 40\% change in the transmission when focused to a 10 $\mu$m RMS spot. This corresponds to a peak fluence from the electron beam on the YAG of $F = 80 $ J/cm$^2$. The signal saturates close to this level. The noise floor of this approach for a single scan (100 shots per time point) is approximately $10^{-3}$. Therefore, even a pulse with a bunch charge of 0.25 pC would still have a signal-to-noise ratio of 5, sufficient to measure temporal overlap. Alternately, a spatially large electron beam with 90 $\mu$m RMS size and 20 pC of bunch charge would produce signal above the noise floor, making it viable for larger electron beams.

\section{Drift and Jitter}

In addition to the characterization of the average spatiotemporal overlap of the pulses, this approach can track the long-term drift of the spatiotemporal overlap. Long-term temporal drift of the electron beam and optical pulses can occur through thermal variation in the amplifier cavities, thermal expansion of the beamlines and optical tables, and drifts in the RF phase relative to the laser timing.  

We track this drift by performing spatial and temporal overlap scans periodically throughout the day and updating the position of the laser beam to the optimal values. The results of these scans figure \ref{fig:Drift} show the long-term spatiotemporal drift of our system with no feedback corrections as measured by the transient YAG approach. For reference, the orange circle in Fig. \ref{fig:Drift}(c) RMS size of the electron beam. Based on these results, the uncorrected spatiotemporal drift will cause the electron and laser beams to fall out of spatiotemporal overlap on the hour timescale, necessitating active feedback of the overlap. 


\section{Conclusion}

We demonstrate a compact diagnostic for spatiotemporal overlap of low-charge electron beams and a counterpropagating laser beam in the context of an ICS x-ray source. The method is compatible with sub-pC bunch charges and sub-$\mu$J laser pulse energies. This technique is broadly applicable to laser/electron beam interactions for a range of laser wavelengths from the mid-infrared to the visible, bounded by the transmission of the scintillator screen and wavelength coverage of commercially available cameras and optical detectors.

\begin{acknowledgments}
This material is based upon work supported by the National Science Foundation under Grant Nos. 2153503 and 2153503. The authors would like to thank David. A. Reis for helpful discussions. 
\end{acknowledgments}

\section*{Author Declarations}
\subsection*{Conflict of Interests}
The authors have no conflicts to disclose.

\section*{Data Availability Statement}
All data and scripts required to produce the figures in this publication are available via GitLab. Data are recorded in the open MATLAB file format, and analysis is available as Jupyter notebooks.

\appendix

\section*{References}
\bibliography{main.bib}
\end{document}